\documentclass[aps,superscriptaddress,amsmath,amssymb,prl,twocolumn,10pt,floatfi
x]{revtex4}
\usepackage[dvips]{graphicx}
\usepackage{epsfig}
\usepackage{amssymb,amsfonts,amsmath}

\begin{document}




\title{Roton softening and supersolidity in Rb spinor condensates}
\author{R. W. Cherng}
\author{E. Demler}
\affiliation{Department of Physics, Harvard University, Cambridge, MA 02138}





\begin{abstract}
Superfluids with a tendency towards periodic crystalline order
have both a phonon and roton like spectrum of collective modes.
The softening of the roton spectrum provides one route
to a supersolid.  We show that roton softening occurs in 
$^{87}$Rb spinor condensates once dipolar interactions and spin
dynamics are
taken into account.  By including the effects of a quasi-two-dimensional
geometry and rapid Larmor precession, we show a dynamical instability 
develops in the collective mode spectrum at finite wavevectors.
We construct phase diagrams showing a variety of instabilities
as a function of the direction of the magnetic field and strength
of the quadratic Zeeman shift.  Our results provide a possible
explanation of current experiments in the Berkeley group 
\textit{Phys. Rev. Lett.} 100:170403 (2008).
\end{abstract}

\maketitle


The experimentally elusive supersolid state has long been of
great interest as an exotic quantum state of matter.
Such a state offers the possibility of periodic crystalline order breaking
translation invariance familiar from solids and
long range phase coherence breaking global gauge invariance 
familiar from superfluids coexisting in the same
material \cite{andreev-69,chester-70,leggett-70,otterlo-95,sengupta-05,scarola-05}.  
Theoretical interest in the supersolid state has a long history
dating back to early studies of superfluidity in $^4$He.
Landau suggested that the excitation spectrum consists of
two parts: a sound like long wavelength mode and
a roton spectrum $\epsilon(p)=\Delta+(p-p_0)^2/2m$ for 
$p\approx p_0$ indicative of a tendency towards crystalline order.
In a system
with both a phonon and roton spectrum,
the softening of the roton gap where $\Delta$ 
approaches zero provides one means of possibly 
realizing a supersolid \cite{kirzhnits-70,schneider-71}.
Recent experiments suggesting the existence of the supersolid
phase in $^4$He are a subject of intense debate 
\cite{chan-04,reppy-06,huse-05,prokofev-07}.

In this paper, we demonstrate roton softening takes place in 
quasi-two-dimensional $F=1$ 
ferromangetic condensates such as ultracold $^{87}$Rb
once dipolar interactions and spin dynamics are taken into account.
Pattern formation due to dipolar interactions is
well known in condensed matter physics \cite{kittel-53}, but spinor condensates
such as $^{87}$Rb present several novel effects without condensed matter
analogs.  In addition to dipolar interactions, the spin degrees of freedom
also experience competing interactions including spin dependent
contact interactions and the quadratic Zeeman shift \cite{dsk-06}.  
Moreover, the spin dynamics of rapid Larmor precession 
and confinement to a two-dimensional geometry play an important role in
modifying the effective dipolar interaction.
Several works have previously analyzed the role of dipolar interactions
in polar molecules 
\cite{lewenstein-00,santos-02,shlyapnikov-03,fischer-06,wang-08}  and 
spinor condensates \cite{meystre-01,pu-06,ueda-06}.  However,
these studies do not take into account the combined effects of dynamical
spin degrees of freedom, rapid Larmor precession, and reduced
dimensionality.

\begin{figure}
\begin{center}
\includegraphics[width=2.25in]{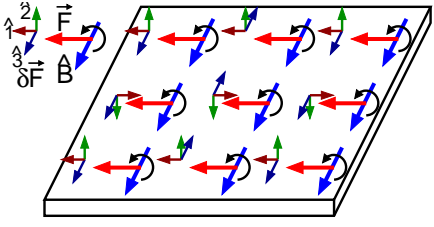}
\end{center}
\caption{
We consider a quasi-two-dimensional spinor condensate with
magnetic field $\hat{B}$ along the $\hat{3}$ direction in the plane.
and uniform magnetization $\vec{F}$.
Larmor precession of $\vec{F}$ about $\hat{B}$ means $\vec{F}$ 
spends half the time along the $\hat{1}$ direction in the plane
and half the time along the $\hat{2}$ direction out of the plane.
Magnetization fluctuations $\delta\vec{F}$ with periodic modulation
can lower the effective dipolar interaction energy and drive an instability
towards a possible supersolid state.
$\delta\vec{F}$ along $\hat{3}$ 
(parallel to $\hat{B}$) favor modulation along
$\hat{1}$ (perpendicular to $\hat{B}$).
In contrast, $\delta\vec{F}$ along $\hat{1}$ or $\hat{2}$
(perpendicular to $\hat{B}$) favor modulation along $\hat{3}$
(parallel to $\hat{B}$).  All components couple to each other
through canonical commutation relations and give rise to a variety
of instabilities towards striped phases modulated along $\hat{1}$
or $\hat{3}$ as well as checkboard phases modudulated along
$\hat{1}$ and $\hat{3}$.
}
\label{fig:instability}
\end{figure}

We show that for an initially uniform ferromagnet, the excitation
spectrum has roton like parts for both spin and density branches
(generally each branch invloves both spin and density degress of
freedom, so we define them by their behavior in the long wavelength
limit) with the roton on the spin branch becoming imaginary at finite
wavevectors.  Macroscopic occupation of such roton excitations should
lead to a state which breaks global gauge invariance, spin rotational
and translational symmetries. Hence such instability suggests the
likely formation of a supersolid phase.
This instability
has a simple physical origin in terms of lowering the classical dipolar 
interaction energy of a uniform ferromagnet through periodic modulation
of the magnetization $\vec{F}$ as shown in Fig. \ref{fig:instability}. 
However, the modulation direction is different for the longitudinal 
component along the direction of the magnetic field $\hat{B}$
versus the perpendicular components because only the latter
precesses around the external magnetic field (see discussion below).
Different components of the spin are conjugate variables
which couple to each other through the commutation relation
$[\vec{F}^i,\vec{F}^j]=i\epsilon^{ijk}\vec{F}^k$, a rich variety of instabilities 
leading to possible striped and checkerboard supersolid phases can arise.  
This simple physical picture also gives a rough estimate
for the length scale $\lambda$ of the instability by equating the kinetic energy
cost with the gain in dipolar interaction energy
$
1/2m\lambda^2+1/2md^2=2\pi g_d n_0/3
$
where $m$ is the mass, $g_d$ gives the strength of dipolar interactions,
$n_0$ is the density, and $d_n$ is the thickness of the condensate.  With typical
experimental parameters $d=2\mu$m and $g_d n_0=10$ Hz this gives an
estimate of $\lambda\approx 10 \mu$ m.  The length scale and structure of
unstable modes agree quantitatively with current experiments on dipolar effects
in spinor condensates \cite{dsk-08}.

\section{\label{sec:results}Results}

\begin{figure}
\begin{center}
\includegraphics[width=2.25in]{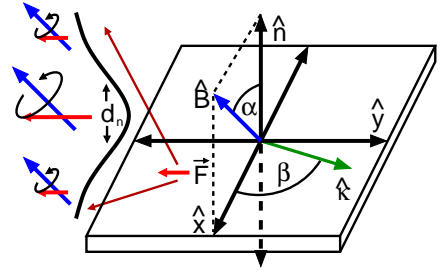}
\end{center}
\caption{
Schematic of experiment with
unit vectors $\hat{x}$, $\hat{y}$ ($\hat{n}$)
in the plane (out of the plane).  
The magnetic field $\hat{B}$ (spiral wavevector $\hat{\kappa}$
makes an angle $\alpha$ ($\beta$) with respect to $\hat{n}$ ($\hat{x}$).
The effective dipolar interaction takes into account Larmor precession
of the magnetization $F$ and the confinement of the condensate along $\hat{b}$ 
to thickness $d_n$.
}
\label{fig:schematic}
\end{figure}

We consider a quasi-two-dimensional spinor condensate as shown
in Fig. \ref{fig:schematic}.  The unit vectors $\hat{x}$, $\hat{y}$
are in the plane while $\hat{n}$ is out of the plane.  A uniform
magnetic field points along $\hat{B}$ in the $\hat{n}$, $\hat{x}$ 
plane at an angle $\alpha$ with respect to $\hat{n}$.  The initial
condensate is either prepared with a uniform magnetization or with
a non-uniform spiral spin texture with magnetization winding
along $\hat{\kappa}$ and wavevector $|\kappa|$.
The Hamiltonian is given by
\begin{align}
\nonumber
\mathcal{H}=&
\int d^3x
\mathbf{\Psi}_{\vec{x}}^\dagger
\left[
-\frac{\nabla^2}{2m}-\mu
+B_0\hat{B}\cdot\vec{F}
+q\left(\hat{B}\cdot\vec{F}\right)^2
\right]
\mathbf{\Psi}_{\vec{x}}\\
\nonumber
&+
\int d^3x
\left[
\frac{g_0}{2}
:\mathbf{\Psi}_{\vec{x}}^\dagger\mathbf{\Psi}_{\vec{x}}
\mathbf{\Psi}_{\vec{x}}^\dagger\mathbf{\Psi}_{\vec{x}}:
+\frac{g_s}{2}
:\mathbf{\Psi}_{\vec{x}}^\dagger\mathbf{\Psi}_{\vec{x}}^*
\mathbf{\Psi}_{\vec{x}}^T\mathbf{\Psi}_{\vec{x}}:
\right]\\
&+\int d^3xd^3x'\frac{g_d}{2} h^{ij}_{3D}(\vec{x}-\vec{x}')
:\mathbf{\Psi}_{\vec{x}}^\dagger \vec{F}^i \mathbf{\Psi}_{\vec{x}}
\mathbf{\Psi}_{\vec{x}'}^\dagger \vec{F}^j \mathbf{\Psi}_{\vec{x}'}:
\label{eq:hamiltonian}
\end{align}
where $:\ :$ denotes normal ordering.
We denote $\mathbf{\Psi}_\alpha$ with $\alpha=1,2,3$ 
as annhilation operators for $F=1$ bosons with mass
$m$ and $\vec{F}$ hyperfine spin operators with
$\vec{F}^i_{jk}=-i\epsilon_{ijk}$.
Throughout, we use a matrix notation with suppressed indices
where $*$, $T$, and $\dagger$ denote the complex conjugate,
transpose, and the conjugate transpose, respectively.
For example, $\mathbf{\Psi}$ ($\mathbf{\Psi}^\dagger$) is a column (row) vector
while $\vec{F}^i$ is a matrix.

The chemical potential $\mu$ is a 
Lagrange multiplier controlling the density
$n_{3D}=\langle \mathbf{\Psi}^\dagger\mathbf{\Psi}\rangle$
and we work with fixed longitudinal magnetization
$n_{3D}f_B=\langle \mathbf{\Psi}^\dagger \hat{B}\cdot \vec{F}\mathbf{\Psi}\rangle$.
The magnetic field induces Larmor precession about $\hat{B}$ at 
a frequency $B_0$ and a quadratic Zeeman shift $q$.
With typical magnetic fields $B$ of zero up to hundreds of mG, 
$q=70$ Hz G$^{-2}$ $B^2$ \cite{dsk-06} ranges from zero to tens of Hz.
AC stark shifts can further tune $q$, in particular to negative values.
A harmonic trapping potential
along $\hat{n}$ confines the condensate to a thickness $d_n$.
We take typical values of $B_0/2\pi=115$ kHz and $d_n=2$ $\mu$m \cite{dsk-08}.

The spin independent and spin dependent contact 
interaction strengths are given by 
$g_0=4\pi\hbar^2a_{0}/m$, $g_s=4\pi\hbar^2(a_{0}-a_{2})/3m$
\cite{ho-98}
in terms of the $s$-wave scattering lengths $a_F$ for two atoms colliding with
total angular momentum $F$.
For $^{87}$Rb, $a_0=101.8 a_B$ and $a_2=100.4 a_B$ where $a_B$ is 
the Bohr radius \cite{verhaar-02}
giving positive $g_s$ and ferromagnetic interactions.
The dipolar interaction strength is given by
$g_d=\mu_0g_F^2\mu_B^2$ where $\mu_0$ is the vacuum permeability,
$g_F$ is the Land\'{e} g-factor, and $\mu_B$ is the Bohr magneton.
The dipolar interaction tensor is given by
\begin{align}
h_{3D}^{ij}(\vec{x})&=|\vec{x}|^{-3}
\left[\delta^{ij}-3\hat{x}^i\hat{x}^j\right],&
h_{3D}^{ij}(\vec{k})&=-\frac{4\pi}{3}
\left[\delta^{ij}-3\hat{k}^i\hat{k}^j\right]
\label{eq:dipolar_3D}
\end{align}
in real and momentum space with the Fourier transform 
regularized as in \textit{Methods:Dipolar Interaction}.
For typical peak three-dimensional densities of $n_{3D}=2.2\times
10^{14}$ cm$^{-3}$ the interaction strengths are
$g_0n_{3D}=1.7$ kHz, $g_sn_{3D}=8$ Hz, and $g_dn_{3D}=10$ Hz 
\cite{dsk-06,dsk-08}.

Notice the clear separation of energy scales for the above
Hamiltonian.  The quadratic Zeeman energy, spin dependent
contact interaction, and dipolar interaction all compete
at the lowest energies on the order of tens of Hz.
The spin independent contact interaction and
harmonic trapping along $\hat{n}$ are both hundreds of Hz
implying the density and
out of plane dynamics are effectively
frozen.
At the highest energies,
Larmor precession occurs at hundreds of kHz implying
one should average over this rapid precession for the
low energy properties.

\subsection{\label{subsec:dipolar}Dipolar Interaction}
As illustrated in Fig. \ref{fig:schematic}, the 
bare dipolar interaction is modified by spatial and time averaging
due to confinement along $\hat{n}$ and Larmor precession, respectively.
Here we briefly summarize the resulting effective dipolar
interaction that emerges with more details
in \textit{Methods:Dipolar Interaction}.


We first consider the effect of confinement along $\hat{n}$.  Since
the thickness of the condensate $d_n$ and the spin healing length
$\xi$ are comparable $d_n, \xi\sim 2$ $\mu$m, we 
assume the condensate is frozen along
$\hat{n}$ and take
\begin{equation}
\mathbf{\Psi}_{\vec{x}}\rightarrow
\sqrt{\rho(x_n)}
\mathbf{\Psi}_{\vec{x}}
\label{eq:trial_wavefn}
\end{equation}
where $\mathbf{\Psi}_{\vec{x}}$ on the left-hand (right-hand) side
is a three-dimensional (two-dimensional) field.
Here $\rho(x_n)$ is the normalized to $\int dx_n \rho(x_n)=1$
and for definiteness we take a gaussian form 
$\rho(x_n)=\exp(-x_n^2/2d_n^2)/\sqrt{2\pi d_n^2}$
with $x_n$ the coordinate along $\hat{n}$.
Integrating over $x_n$ in Eq. \ref{eq:hamiltonian} gives
a two-dimensional Hamiltonian with
$d^3x\rightarrow d^2x$, 
$g_0\rightarrow g_0\rho(0)C$,
$g_s\rightarrow g_s\rho(0)C$,
$g_d\rightarrow g_d\rho(0)C$
where the constant $C=1/\sqrt{2}$ is 
determined by normalization.
However, the dipolar interaction tensor 
$h^{ij}_{3D}(\Delta x)\rightarrow h^{ij}_{2D}(\Delta x)$
is modified appreciably.

In addition to confinement along $\hat{n}$, rapid 
Larmor precession also modifies the dipolar interaction tensor.
In experiments, the strong uniform component of the magnetic field 
causes precession of magnetization along $\hat{B}$ while a weak gradient
along $\hat{\kappa}$ induces a spiral order of the magnetization \cite{dsk-08}.
We take this into account by going to a co-moving frame via
the unitary transformation
\begin{align}
\mathbf{\Psi}_x&\rightarrow R(t,\vec{x})\mathbf{\Psi}_x,&
R(t,\vec{x})&=\exp\left[i\theta(t,\vec{x})\hat{B}\cdot \vec{F}\right]
\label{eq:rotating_frame}
\end{align}
with $\theta(t,\vec{x})=-B_0 t+|\kappa|\hat{\kappa}\cdot\vec{x}$
where $B_0$ is the uniform component of the magnetic field and 
$|\kappa|$ is the spiral wavevector.
Applying this unitary transformation simply yields the substitutions
$B_0\rightarrow-i|\kappa|/m\hat{\kappa}\cdot\nabla$,
$q\rightarrow q+|\kappa|^2/2m$
which arise from the Berry's phase and kinetic energy terms.
Time averaging over the rapid precession then yields 
a modified dipolar interaction tensor
$h^{ij}_{2D}(\Delta x)\rightarrow \bar{h}^{ij}_{2D}(\Delta x)$.

We focus on the resulting two-dimensional time-averaged interaction in momentum
space given by 
\begin{align}
\label{eq:dipolar_effective}
\bar{h}_{2D}^{ij}(\vec{k})&=
-\frac{4\pi}{3}
\left[
\bar{h}_{2D}^B(\vec{k})P^B_{ij}
+\bar{h}^\perp_{2D}(\vec{k})P^{\perp}_{ij}
+\bar{h}^\times_{2D}(\vec{k})\hat{B}\cdot \vec{F}_{ij}
\right]
\end{align}
with the momentum independent projection operators
\begin{align}
P^B_{ij}&=\hat{B}^i\hat{B}^j,&
P^\perp_{ij}&=\delta^{ij}-\hat{B}^i\hat{B}^j
\end{align}
and recall $\vec{F}^k_{ij}=-i\epsilon_{ijk}$
while the functions 
$\bar{h}^B_{2D}(\vec{k})$, $\bar{h}^\perp_{2D}(\vec{k})$, 
$\bar{h}^\times_{2D}(\vec{k})$ carry the momentum
dependence and are given explicitly in 
Eqs. \ref{eq:hb}, \ref{eq:hp}, \ref{eq:hx} of
\textit{Methods:Dipolar Interaction}.
Notice the spin dependent part carrying the $i,j$
indices only depends on $\hat{B}$ and not $\vec{k}$.  
Essentially, time-averaging over the fast Larmor precession selects a preferred
direction in spin space along $\hat{B}$.
Compare this to the bare dipolar interaction of Eq. \ref{eq:dipolar_3D}.
where the spin and momentum dependence are not separable.  
In real space, the bare dipolar interaction
(without averaging over Larmor precession)
favors spins aligned head-to-tail and
anti-aligned side-by-side.
In particular,
small distortions of a mean-field condensate with uniform magnetization 
in the plane are energetically unfavorable since they destroy the
favorable head-to-tail order already present.

These considerations change once rapid Larmor precession is
included.  Precession causes the energetically favorable
head-to-tail order to rotate into the energetically unfavorable
side-by-side order half of the time.  Heuristically, this
gives rise the instabilities of the effective dipolar interaction.
Fig. \ref{fig:instability} illustrates the
fluctuations $\delta \vec{F}$ for the magnetization $\vec{F}$
that lower the effective dipolar energy with $\hat{B}$ in the plane.
$\delta\vec{F}$ parallel to $\hat{B}$ favors modulation 
perpendicular to $\hat{B}$.  This is as expected even for the
bare dipolar interaction which favors head-to-tail alignment
(uniform parallel to $\hat{B}$) and side-by-side anti-alignment
(modulation perpendicular to $\hat{B}$).  However, both 
components of $\delta\vec{F}$ perpendicular to $\hat{B}$
favor modulation parallel to $\hat{B}$.  For $\delta\vec{F}$ in the plane
and perpendicular to $\hat{B}$, this can be understood in terms of the
bare dipolar interaction favoring head-to-tail alignment
(uniform perpendicular to $\hat{B}$) and side-by-side anti-alignment
(modulation parallel to $\hat{B}$).  Larmor precession affects
$\delta\vec{F}$ out of the plane and perpendicular to $\hat{B}$ the
most as precession rotates this component into the plane half the time.
This leads to the same type of behavior as $\delta\vec{F}$
in the plane and perpendicular to $\hat{B}$:
uniform perpendicular
to $\hat{B}$ and modulation parallel to $\hat{B}$.  Compare this to
arguments appealing to the
bare dipolar interaction for this component which suggest modulation
along both directions since
$\delta\vec{F}$ out of the plane gives
the energetically unfavorable aligned side-by-side arrangement.

\subsection{\label{subsec:cm}Collective Modes}
We now turn to quantitative analysis of the collective mode spectrum focusing 
on the two-dimensional time-averaged case.
Starting from mean field solutions with uniform magnetization,  
we study the collective mode spectrum describing its small fluctuations.
For details see 
\textit{Methods:Collective Modes}.
Recall
we transformed to a frame co-moving with possible spiral order
so that co-moving frame uniform states
describe both lab frame uniform and spiral states.

We take $\hat{B}$ as the quantization axis and
parametrize
\begin{equation}
\mathbf{\Psi}_{\vec{x}}=\sqrt{n_{\vec{x}}}
\begin{bmatrix}
ie^{i\eta_{\vec{x}}+i\nu_{\vec{x}}}
\cos(\phi_{\vec{x}}+i\chi_{\vec{x}})\frac{\sin(\rho_{\vec{x}})}
{\sqrt{\cosh(2\chi_{\vec{x}})}}\\
ie^{i\eta_{\vec{x}}+i\nu_{\vec{x}}}
\sin(\phi_{\vec{x}}+i\chi_{\vec{x}})\frac{\sin(\rho_{\vec{x}})}
{\sqrt{\cosh(2\chi_{\vec{x}})}}\\
e^{i\eta_{\vec{x}}}\cos(\rho_{\vec{x}})
\end{bmatrix}
\label{eq:mean_field_parametrization}
\end{equation}
with $n$ the two-dimensional density, $\eta$ the global phase,
$\rho$, $\chi$, $\nu$  controlling the magnitude of the magnetization, and 
$\phi$ the orientation of the transverse magnetization.
We take $\mathbf{\Psi}_{\vec{x}}=\mathbf{\Psi}$
independent of $\vec{x}$ and 
show in \textit{Methods:Collective Modes}
these mean-field states only depend on
\begin{align}
Q&=\frac{q}
{2g_\perp n_{3D}C},&g_\perp=g_s-g_d \bar{h}^{\perp}_{2D}(0)
\label{eq:mean_field_parameters}
\end{align}
where we use $n_{3D}=n\rho(0)$.

To study collective modes, we take 
$\mathbf{\Psi}_{\vec{x}}=\mathbf{\Psi}+\delta\mathbf{\Psi}_{\vec{x}}$
and find the linearized equations of motion for the fluctuations
$\delta\mathbf{\Psi}_{\vec{k}}$ given by
\begin{equation}
i\partial_t
\begin{bmatrix}
\delta\mathbf{\Psi}_{\vec{k}}\\
\delta\mathbf{\Psi}_{-\vec{k}}^*
\end{bmatrix}
=\begin{bmatrix}
M_{\vec{k}}&N_{\vec{k}}\\
-N_{-\vec{k}}^*&-M_{-\vec{k}}^*
\end{bmatrix}
\begin{bmatrix}
\delta\mathbf{\Psi}_{\vec{k}}\\
\delta\mathbf{\Psi}_{-\vec{k}}^*
\end{bmatrix}
\label{eq:collective_modes_eom}
\end{equation}
with $M_{\vec{k}}$ and $N_{\vec{k}}$ given in 
\textit{Methods:Collective Modes}.
The ansatz 
$\delta\mathbf{\Psi}_{\vec{k}}(t)\sim e^{i\omega_{\vec{k}} t}$
gives an eigenvalue equation for the excitation energies
$\omega_{\vec{k}}$.
Analysis of $\omega_{\vec{k}}$ gives the spectrum 
for small fluctuations above the mean field solution.
In particular, imaginary $\omega_{\vec{k}}$ indicates a dynamical instability
where such fluctuations grow exponentially.

The above collective mode analysis describes the general
case for arbitrary $f_B,|\kappa|\ne 0$.  
Here we focus on the analytically tractable case of
zero longitudinal magnetization and no spiral order 
$f_B,|\kappa|=0$ and later discuss the effects of finite 
$f_B,|\kappa|$.
Instead of the three component complex field $\mathbf{\Psi}_{\vec{x}}$,
it will be convenient to use the six component real field
\begin{equation}
\mathbf{\Phi}_{\vec{x}}=
\begin{bmatrix}
n_{\vec{x}}&\eta_{\vec{x}}&\rho_{\vec{x}}&\nu_{\vec{x}}&\phi_{\vec{x}}&\chi_{\vec{x}}
\end{bmatrix}^T
\end{equation}
where Eq. \ref{eq:mean_field_parametrization}
defines $\mathbf{\Psi}_{\vec{x}}$ as a function of $\mathbf{\Phi}_{\vec{x}}$.
For the collective mode analysis, we take
\begin{equation}
\delta\mathbf{\Psi}_{\vec{x}}=
\frac{\partial \mathbf{\Psi}}{\partial \mathbf{\Phi}}
\delta\mathbf{\Phi}_{\vec{x}}
\end{equation}
with the derivative evaluated at the mean-field parameters
and derive equations of motion for
$\delta \mathbf{\Phi}_{\vec{k}}$ from Eq. \ref{eq:collective_modes_eom}.
The $\phi$ and $\chi$ modes decouple from
the $n$, $\eta$, $\rho$, and $\nu$ modes.
Since $\phi$ controls the orientation of the transverse
magnetization and $\chi$ controls the magnitude of the longitudinal
magnetization, they form a conjugate pair of variables which we denote
as the spin mode.
The other four degrees of freedom form two pairs of conjugate
variables which we denote as the charge and magnetization modes.

\begin{figure}
\begin{center}
\includegraphics[width=2.25in]{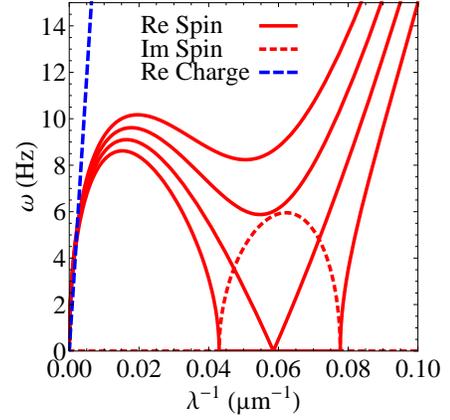}
\end{center}
\caption{
Collective mode spectrum for $\alpha=0.24\pi$, $\theta=\pi/2$ and 
$q=q_c+\delta q$ with $q_c=-0.87$ and $\delta q=1.0, 0.5, 0.0, -0.5$ Hz
from top to bottom.
}
\label{fig:roton}
\end{figure}

At long wavelengths, we find for the spin, charge, and 
magnetization modes
\begin{align}
\nonumber
\omega^s_k=&
\sqrt{\left[1+Q\right]
\left[(Q-1)g_s+(3-Q)g_\perp(\alpha)\right]g_d n_{3D}^2C}\\
&\times\sqrt{\left[\cos^2(\alpha)-\sin^2(\alpha)\cos^2(\theta)\right]}\ 
\sqrt{\frac{\pi k}{\rho(0)}}\\
\omega^c_k=&
\sqrt{2\left[g_0+g_s-g_\perp(\alpha)\right]n_{3D}C}\ \sqrt{\frac{k^2}{2m}}\\
\omega^m_k=&
2|1-Q^2||g_\perp(\alpha)|n_{3D}C
\end{align}
where $\alpha$ ($\theta$ )is the angle between $\hat{B}$ and $\hat{n}$,
($\hat{k}$ and $\hat{x}$).
In the above, we write out the explicit $\alpha$ dependence of
Eq. \ref{eq:mean_field_parameters} as
\begin{equation}
g_\perp(\alpha)=g_s-\frac{2\pi}{3}g_d\left[1-3\cos^2(\alpha)\right]
\end{equation}
The spin mode scales with $\sqrt{k}$ and is highly anisotropic
due to the long-ranged and anisotropic nature of dipolar 
interactions.  The spin mode can develop a roton 
minimum indicating a tendency towards crystalline order
and has a strong dependence on $\alpha$.  
The charge mode scales with $k$ and describes 
phonon excitations of the superfluid.  Notice the superfluid velocity
depends primarily on the spin-independent contact interaction
$g_0$.  The magnetization mode
is gapped and describes fluctuations in the magntiude of
$\vec{F}$.  Notice the gap can vanish as a function
of $Q$ and $\alpha$. 

We plot a representative collective
mode spectrum in Fig. \ref{fig:roton} illustrating
the spin and charge modes for $\alpha=0.24 \pi$, $\theta=\pi/2$ and
$q$ near $q_c=-0.87$ Hz.  Notice the appearance of a roton
minimum and the softening of the roton gap as $q$ approaches
$q_c$.  When $q$ is below $q_c$, the spin mode becomes imaginary
at finite wavevector indicating a dynamical instability
towards periodic crystalline order.

\begin{figure*}
\begin{center}
\includegraphics[height=2.5in]{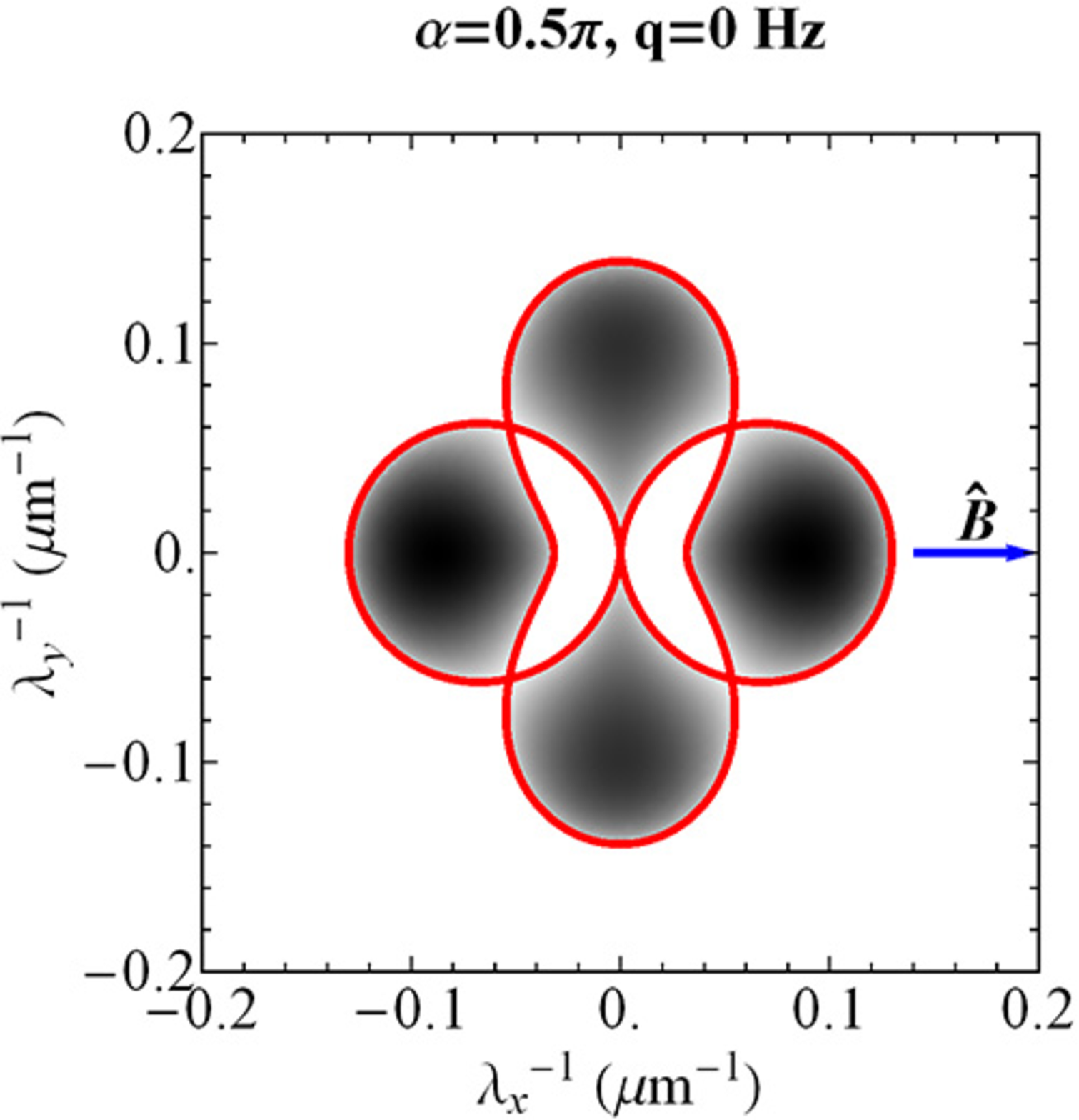}
\includegraphics[height=2.5in]{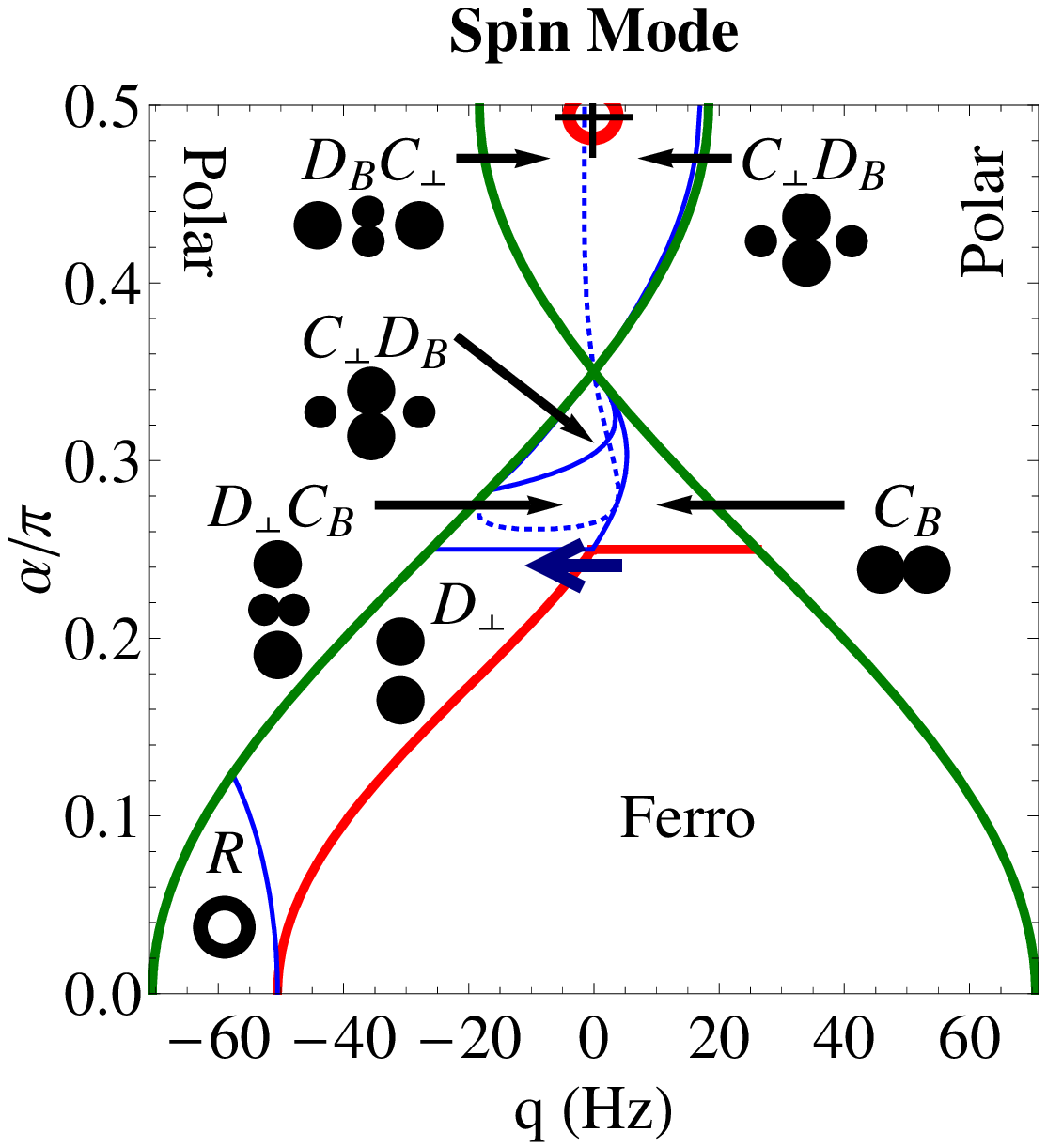}
\includegraphics[height=2.5in]{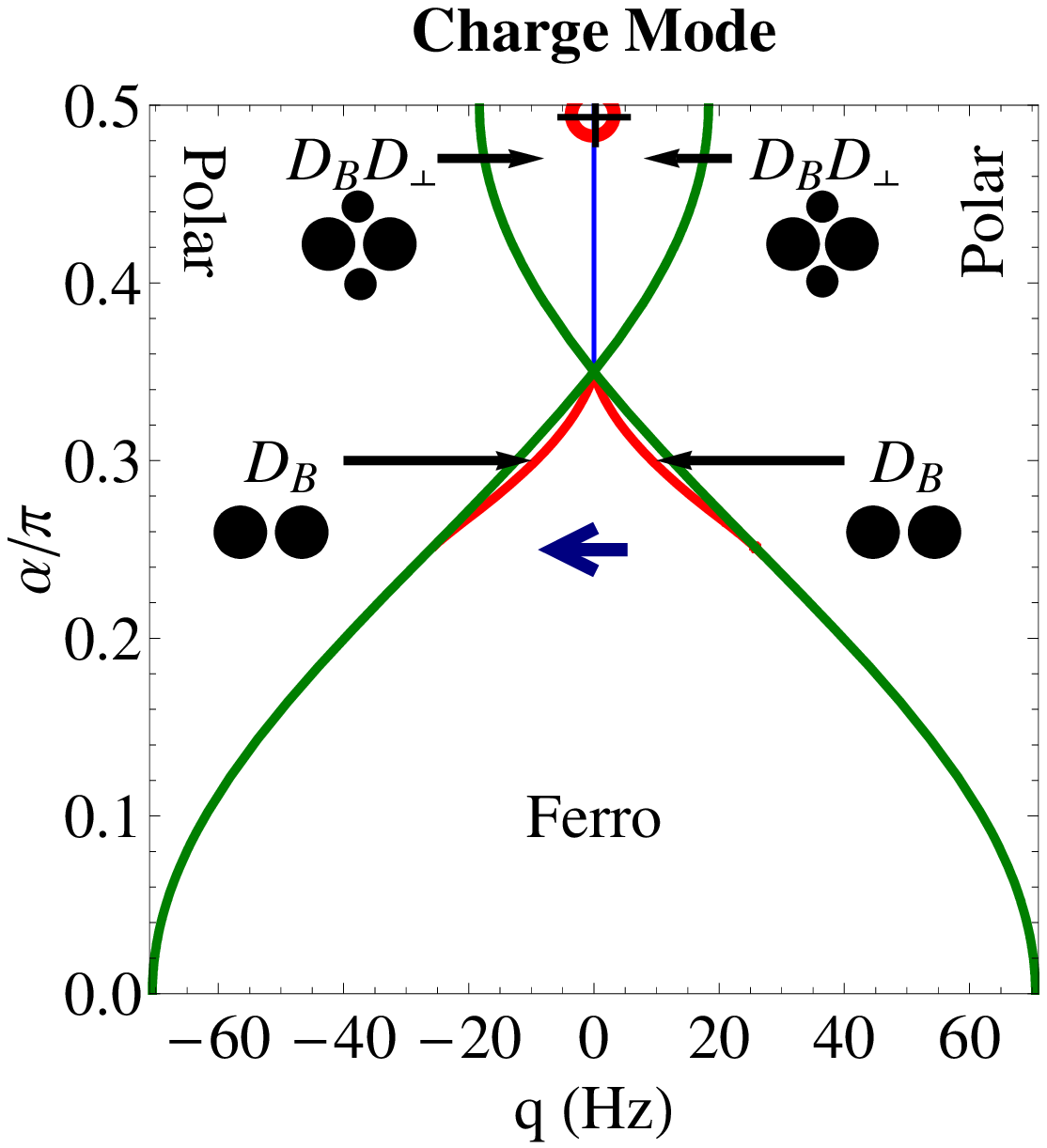}
\end{center}
\caption{
Spin mode in momentum space (left) illustrating regions of unstable modes
with shading indicating magnitude of
$\text{Im}\ \omega_{\vec{k}}$ and $\hat{B}$ the magnetic field 
for current experimental parameters $\alpha=\pi/2$, $q=0$ \cite{dsk-08}. 
The red marker indicates these parameters in the spin and
charge mode phase diagrams (middle and left) which have
similar plots illustrating regions of unstable modes in various phases
(see text for description).  The blue arrow indicates parameters
for Fig. \ref{fig:roton}.
}
\label{fig:phase_diagrams}
\end{figure*}

We then analyze the imaginary part of the collective mode spectrum
in momentum space as in the left of Fig.
\ref{fig:phase_diagrams}.  This indicates the structure
of instabilities and we construct the phase diagrams
for the spin and charge mode in Fig. \ref{fig:phase_diagrams}.
The uniform ferromagnet mean-field solution exists within the green lines
and the polar state with no magnetization is outside.
Dynamical instabilities above the red line imply the mean-field
state is unstable to small fluctuations. 
In general, spin mode instabilities are stronger with larger 
$\text{Im}\ \omega_{\vec{k}}$ than charge mode instabilities
except near the boundaries to the polar state where they are comparable.

Notice the boundaries to the polar state intersect 
near $\alpha_c=0.35 \pi$ indicating the uniform ferromagnet is
not stable for any $q$ even at the mean-field level.  This is
analgous to the magic angle effect familiar from NMR \cite{nmr-83}.  At 
this magic angle, $g_\perp(\alpha_c)=0$ and rapid precession
of the spins cancels out the combined effect of dipolar and
spin dependent contact interactions which stabilize the ferromagnet.

In the phase diagrams,
$D_\alpha$ ($C_\alpha$) denotes regions of unstable modes
disconnected from (connected to) the origin $k=0$ and $R$
indicates a ring of unstable modes.  There are several
trends to notice in the phase digrams.  The uniform ferromagnet 
is stable only for $\alpha<\pi/4$ away from the negative $q$ boundary.
For $\alpha<\pi/4$ near the negative $q$ boundary the spin mode is unstable
along the direction perpendicular to $\hat{B}$.
This suggests an instability towards a striped phase
modulated along this direction.  From 
Fig. \ref{fig:instability}, we see it is 
driven by transverse fluctuations $\delta \vec{F}$ of the magnetization
$\vec{F}$ which modulation perpendicular to $\hat{B}$.
There is also an area of ring instabilities near the lower left hand corner.  
Generally, modes perpendicular to $\hat{B}$ are more unstable than modes 
parallel to $\hat{B}$.

For $\pi/4<\alpha<\alpha_c$, the primary instability is in the spin mode
although the charge mode is also unstable in limited regions near
the polar state.  Again, transverse fluctuations 
$\delta \vec{F}$ drive the spin mode instability towards modulation
perpendicular to $\hat{B}$
near the negative $q$ boundary to the polar state.
However, longitudinal fluctuations $\delta \vec{F}$
drive the spin mode instability towards modulation
parallel to $\hat{B}$ near the positive $q$ boundary.  
This suggests striped phases
perpendicular and parallel to $\hat{B}$ near the negative and
positive $q$ boundaries, respectively.
In between these boundaries, both transverse and longitudinal fluctuations
$\delta \vec{F}$ are important.
This indicates a tendency towards
a checkerboard phase modulated along both directions.
Above the magic angle $\alpha_c<\alpha$, both types of instabilities
are important throughout indicating a tendency towards
checkerboard phases.

\begin{figure}
\begin{center}
\includegraphics[width=3.4in]{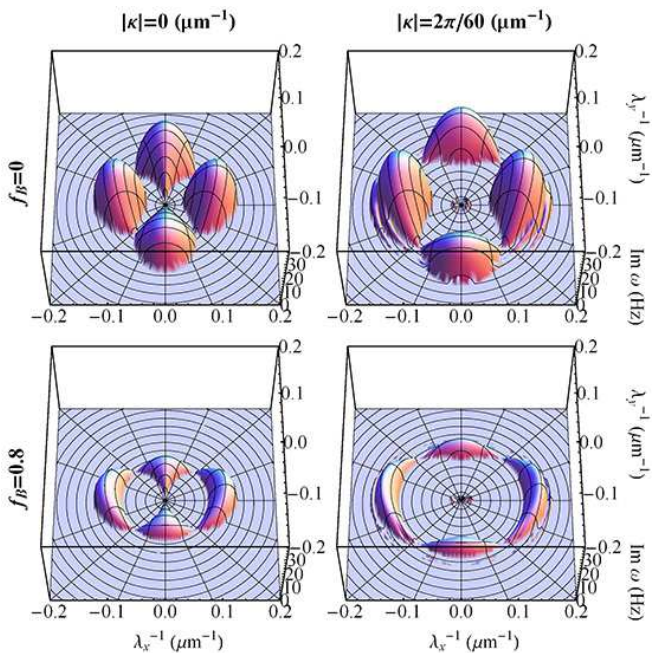}
\end{center}
\caption{
Imaginary part of the spin mode in momentum space
with and without spiral order (right, left) as well as
with and without longitudinal magnetization (bottom, top).
Here we take current experimental parameters of
$\alpha=\pi/2$, $q=0$ Hz, $\beta=0$ \cite{dsk-08}.
}
\label{fig:omega}
\end{figure}

The above analysis for $f_B,|\kappa|=0$ also holds qualitatively for general 
parameters $f_B,|\kappa|\ne 0$.  As an illustration, we plot
in Fig. \ref{fig:omega}
the imaginary part of $\omega_{\vec{k}}$ for 
the experimentally relevant parameters $\alpha=\pi/2$, $q=0$,
$\beta=0$ in four cases: 
$f_B=0$ and $|\kappa|=0$ (top left),
$f_B=0.8$ and $|\kappa|=\kappa_{max}$ (bottom left),
$f_B=0$ and $|\kappa|=0$ (top right),
$f_B=0.8$ and $|\kappa|=\kappa_{max}$ (bottom right).
Here $\kappa_{max}=2\pi/60$ $\mu$m$^{-1}$ is the largest
spiral wavevector (tightest winding) obtainable in current
experiments \cite{dsk-08}.  In general, increasing $f_B$ suppreses
the dynamical instability by decreasing
of $\text{Im}\ \omega_{\vec{k}}$.  Increasing $|\kappa|$ introduces 
additional instabilities with a length scale set by $2\pi/|\kappa|$
which is typically 60 $\mu$m or greater.
They are well-separated from 
instabilities due to dipolar interactions
which are typically at a length scale of 10 $\mu$m.
Experimentally, the instability also appears to occur on shorter or longer timescales
depending on whether the length scale of the spiral winding is smaller or larger,
repspectively, than the largest length scale of the condensate. 
We do not find a strong dependence of the instability timescale on the
wavelength of the spiral winding which was reported in \cite{dsk-08}.
One possibility for this discrepancy is that in experiments, spiral states
are prepared dynamically by applying a strong gradient of the magnetic
field.  Winding the spins dynamically may introduce noise and fluctuations
into the system and facilitate the development of fragementation.

\section{\label{sec:discussion}Discussion} 


Our analysis shows the intriguing complexity of excitation spectrum
of spinor condensates when dipolar interactions are taken into
account. In particular we demonstrated that softening of the roton gap
is associated with transverse fluctuations of the magnetization
This lowers the dipolar interaction energy and such fluctuations
are distinct from phonon excitations of the superfluid.  
Earlier studies on dipolar effects in polar molecules 
treating the spin degrees of freedom as frozen 
\cite{lewenstein-00,santos-02,shlyapnikov-03,fischer-06,wang-08} 
obtained an excitation spectrum with one branch mixing the phonon and roton
parts.

Dynamical spin degrees of freedom have been considered before in studies
of ground state properties of spinor condensates but only with the
bare dipolar interactions.
Our analysis demonstrates rapid Larmor precession and reduced dimensionality
significantly modify the effective dipolar interaction.  Using this
effective interaction, we have shown
uniform mean-field condensates have instabilities towards a possible
supersolid with stripe or checkerboard crystalline order.

Our work opens several new directions in the study of spinor
condensates. We demonstrated that the uniform phase is unstable toward
spin modulation for a whole range of wavevectors. However competition
between unstable modes and the resulting stable state still need to be
explored. Possible candidate phases include both states with small
modulation and states with spiral spin winding. Another interesting
question to be explored is the interplay of Bose condensation and spin
modulation in the presense of thermal fluctuations. At the mean-field
level spin modulation should appear simultaneously with the appearance
of the condensate. However spin modulation breaks translational
symmetry and is a distinct symmetry breaking from the Bose
condensation. Hence apriori the two may have different transition
temperatures.

The main significance of our work is developing a microscopic theory
which successfully explains the checkerboard pattern observed in
recent experiments of the Berkeley group \cite{dsk-08}. The Fourier transform
of the magnetization observed in experiments had a distinct cross-like
structure with a lengthscale of 10 $\mu$m which is in excellent
agreement with the results shown in Figs. \ref{fig:phase_diagrams}
and  \ref{fig:omega}. Previous theoretical
studies \cite{lamacraft-07b,cherng-08}
have addressed these experiments without taking into
account the precession averaged dipolar interaction and have not
explained the observed pattern of spin fragmentation. Although we
analyzed the simplest geometry of an infinite two dimensional layer,
our results are in good agreement with the current experimental
data. We expect however that the two dimensional model may not capture
some effects which may be present in real systems such as pinning of
the instabilities by the in plane trapping potential.

In conclusion, we have considered the combined effects of 
dynamical spin degrees of freedom,
Larmor precession, and reduced dimensionality on the collective mode
spectrum of spinor condensates with dipolar interactions.
Starting from a mean-field state with uniform or spiral
magnetization, we demonstrated the presence of a phonon and 
roton spectrum of collective modes.  Softening of the roton gap
suggest instabilities towards a possible supersolid state
with stripe or checkerboard periodic crystalline order.

\section{Methods}
\subsection{\label{subsec:methods_dipolar}Dipolar Interaction}
Here we outline the derivation of the effective
two-dimensional time-averaged dipolar interaction.
The effect of the transformation in Eq. 
\ref{eq:rotating_frame} is clearest in the bosonic coherent
state path integral formalism for 
$\mathcal{U}$ the evolution operator
\begin{align}
\mathcal{U}&=
\int\mathcal{D}\mathbf{\Psi}^\dagger\mathcal{D}\mathbf{\Psi}
e^{\int dt i\mathcal{L}},&
\mathcal{L}
=i\int d^3x \mathbf{\Psi}^\dagger\partial_t\mathbf{\Psi}
-\mathcal{H}
\end{align}
where $\mathcal{H}$ is given in Eq. \ref{eq:hamiltonian}.
Only the time derivative,
spatial derivative and dipolar interaction tensor transform non-trivially.
The transformed time and spatial derivatives
\begin{align}
\nonumber
\partial_t&\rightarrow \partial_t+B_0 \hat{B}\cdot \vec{F},&\\
-\frac{\nabla^2}{2m}&\rightarrow
-\frac{\nabla^2}{2m}
-i\frac{|\kappa|\hat{\kappa}\cdot\nabla}{m}\hat{B}\cdot \vec{F}
+\frac{|\kappa|^2}{2m}\left(\hat{B}\cdot \vec{F}\right)^2
\end{align}
can be absorbed into Hamiltonian as modified linear and quadratic Zeeman
shifts.

For the dipolar interaction tensor, the transformation needs to
be expressed as one acting on $h^{ij}_{3D}(\vec{x})$ instead of
$\mathbf{\Psi}_{\vec{x}}$.
Recall $\vec{F}^i_{jk}$ are $SO(3)$ Clebsch-Gordon coefficients projecting the tensor
product of two spin-1 representations (lower indices) onto the
spin-1 component (upper index).  In particular,
\begin{equation}
R^T_{jj'}\vec{F}^i_{j'k'}R_{k'k}=R_{ii'}\vec{F}^{i'}_{jk}
\label{eq:f_transformation}
\end{equation}
for an arbitrary $SO(3)$ rotation $R$
implying $R$ acting on the lower two indices is equivalent to $R$ acting 
on the one upper index.
This gives
\begin{equation}
\bar{h}^{ij}_{3D}(\vec{x}-\vec{x}')=
\int_{-\pi/B_0}^{+\pi/B_0}B_0dt
\left[
R(t,\vec{x})^T_{ii'}
h_{3D}^{i'j'}(\vec{x}-\vec{x}')
R(t,\vec{x}')_{j'j}
\right]
\end{equation}
where the bar denotes time-averaging and
the explicit dependence on $t$ and 
$\vec{x}+\vec{x}'$ is removed as a result.
Here $h_{3D}^{ij}(\vec{x})$ is given in 
Eq. \ref{eq:dipolar_3D}.
The short-distance singularity $|\vec{x}|^{-3}$ in the 
Fourier transform
\begin{equation}
\bar{h}^{ij}_{3D}(\vec{k})=
\int d^3x e^{-i\vec{k}\cdot\vec{x}}\bar{h}^{ij}_{3D}(\vec{x})
\end{equation}
is regularized with the prescription
\begin{equation}
|\vec{x}|^{-3}_{reg}=
\begin{cases}
0&|\vec{x}|\le b\\
|\vec{x}|^{-3}&|\vec{x}|>b
\label{eq:dipolar_regularized}
\end{cases}
\end{equation}
and taking $b\rightarrow0$ at the end.

To take into account confinement along $\hat{n}$ to 
thickness $d_n$, we 
consider a general trial wavefunction of the form in 
Eq. \ref{eq:trial_wavefn} with
\begin{align}
\rho(x_n)&=\frac{1}{d_n}f\left(\frac{x_n}{d_n}\right)
\end{align}
where $x_n$ is the coordinate
along $\hat{n}$ and $f(x)$ is normalized to $\int dx f(x)=1$.
This results in
\begin{equation}
\bar{h}_{2D}^{ij}(\vec{k})=
\frac{\int dk_n dx dx' \bar{h}_{3D}^{ij}(\vec{k},k_n)
e^{ik_nd_n (x-x')}f(x)f(x')}
{\int dx f^2(x)}
\end{equation}
with $\vec{k}$ ($k_n$) the two-dimensional (one-dimensional)
coordinate perpendicular (parallel) to $\hat{n}$.
We obtain Eq. \ref{eq:dipolar_effective} with
\begin{align}
\label{eq:hb}
\bar{h}_{2D}^B(\vec{k})&=
1-3g(\vec{k},0)\\
\label{eq:hp}
\bar{h}_{2D}^\perp(\vec{k})&=
-\frac{1}{2}
+\frac{3}{4}g(\vec{k},+|\kappa|\hat{\kappa})
+\frac{3}{4}g(\vec{k},-|\kappa|\hat{\kappa})\\
\label{eq:hx}
\bar{h}_{2D}^\times(\vec{k})&=
-\frac{3}{4}g(\vec{k},+|\kappa|\hat{\kappa})
+\frac{3}{4}g(\vec{k},-|\kappa|\hat{\kappa})
\end{align}
and the function $g(\vec{u},\vec{v})$ given by
\begin{align}
\nonumber
g(\vec{u},\vec{v})=&
\left(\hat{B}\cdot\frac{\vec{u}+\vec{v}}{|\vec{u}+\vec{v}|}\right)^2
w(|\vec{u}+\vec{v}|d_n)
-2\int^{|\vec{u}+\vec{v}|d_n}_{|\vec{u}|d_n}\frac{dq}{q}w(q)\\
&+\left(\hat{B}\cdot\hat{n}\right)^2 \left[1-w(|\vec{u}+\vec{v}|d_n)\right]
\end{align}
and the following function
\begin{align}
w(q)&=\frac{q\int dx dx' e^{-q|x-x'|}f(x)f(x')}
{2\int dx f^2(x)}
\end{align}
the only quantity dependent on $f(x)$.
The asymptotic behavior of $w(q)$ is
\begin{equation}
w(q)=\begin{cases}
\frac{1}{2D_0}q-\frac{D_{+1}}{2D_0}q^2&q\ll 1\\
1-\frac{D_{-2}}{D_0}q^{-2}&q\gg 1
\end{cases}
\end{equation}
with $D_0=\int dx f^2(x)$, 
$D_{+n}=\int dx dx' f(x) f(x') |x-x'|^n$,
$D_{-n}=\int dx f(x)(-i\partial_x)^n f(x)$.
For a gaussian trial wavefunction
\begin{align}
f(x)&=\frac{e^{-x^2/2}}{\sqrt{2\pi}},&
w(q)&=2q\int_0^{\infty}dk e^{-(k^2+2kq)}
\end{align}
and we stress the qualitative behavior of $w(q)$ is 
rather insensitive to the detailed form of $f(x)$.

\subsection{\label{subsec:methods_cm}Collective Modes}

Next we outline the mean-field and collective mode analysis.
Taking $\mathbf{\Psi}_{\vec{x}}=\mathbf{\Psi}$
with all quantities independent of $\vec{x}$ in
Eq. \ref{eq:mean_field_parametrization}
and extremize the free energy
at fixed density $n$ and longitudinal magnetization $f_B$.
The free energy is independent of $\eta$ and $\phi$ and we find two classes
of solutions
\begin{align}
\nu&=0,&
Q\tau^3+(1-Q)\tau&=f_B&
\\
\nu&=\frac{\pi}{2},&
Q\tau^{-3}+(1-Q)\tau^{-1}&=f_B
\label{eq:mean_field_solutions}
\end{align}
with $\sin(\rho)^2=\frac{f_B}{2}\left(\tau+\tau^{-1}\right)$.
The constraints $0\le\sin(\rho)^2\le1$ and $-1\le\tau\le 1$
select the unique root for the cubic equations
\begin{align}
\tau&=\begin{cases}
2\mathcal{Q}\sinh\left[\frac{1}{3}\text{arcsinh}
\left(\frac{f_B}{2Q\mathcal{Q}^3}\right)\right],&
Q<1\\
f_B^{1/3}&Q=1\\
2\mathcal{Q}\text{sign}(f_B)\cosh\left[\frac{1}{3}\text{arccosh}
\left(\frac{|f_B|}{2Q\mathcal{Q}^3}\right)\right],&
Q>1
\end{cases}\\
\tau^{-1}&=
2\mathcal{Q}\text{sign}(f_B)
\cosh\left[\frac{1}{3}\text{arccosh}
\left(\frac{|f_B|}{2Q\mathcal{Q}^3}\right)\right]
\label{eq:mean_field_roots}
\end{align}
with $\mathcal{Q}=\sqrt{|1-Q|/3Q}$ and $Q$ given by
Eq. \ref{eq:mean_field_parameters}
for the $\nu=0,\pi/2$ solutions, respectively.
For $g_\perp(\alpha)>0$ or $\alpha>\alpha_c$
($g_\perp(\alpha)<0$ or $\alpha<\alpha_c$),
the $\nu=0$ ($\nu=\pi/2$) solution has a lower 
mean-field free energy.  However, we stress both
solutions are stationary points for all $\alpha$.

The above constraints also imply the $\nu=0$ solution 
exists for
$Q>-\frac{1}{2}\left(1+\sqrt{1-f_b^2}\right)$ while the
$\nu=\pi/2$ solution exists for
$Q<-\frac{1}{2}\left(1-\sqrt{1-f_b^2}\right)$.
In general, both types of solutions support transverse
magnetization except the fully polarized state
along $f_B=\pm 1$ and the polar state along
$Q>1$, $f_B=0$ ($Q<0$, $f_B=0$) for $\nu=0$ ($\nu=\pi/2$).
We focus on the $\eta=0$ solution for $-1<Q<+1$ as it support a finite
transverse magnetization at $f_B=0$ as observed in experiment.

Turning to the collective mode analysis, the linearized equations
of motion are given by Eq. \ref{eq:collective_modes_eom}
with 
\begin{align}
\nonumber
M_{\vec{k}}=&
\frac{|\vec{k}|^2}{2m}-\mu
-\left(p-\frac{|\kappa|\hat{\kappa}\cdot \vec{k}}{m}\right)\hat{B}\cdot\vec{F}+
\left(q+\frac{|\kappa|^2}{2m}\right)\left(\hat{B}\cdot\vec{F}\right)^2\\
\nonumber
&+g_0\rho(0)C\mathbf{\Psi}^\dagger\mathbf{\Psi}
+g_0\rho(0)C\mathbf{\Psi}\mathbf{\Psi}^\dagger
+2g_s\rho(0)C\mathbf{\Psi}^*\mathbf{\Psi}^T\\
\nonumber
&+g_d\rho(0)C\bar{h}^{ij}_{2D}(0)
\text{Tr}\left[\vec{F}^i \mathbf{\Psi}\mathbf{\Psi}^\dagger\right]\vec{F}^j\\
&+g_d\rho(0)C \bar{h}^{ij}_{2D}(\vec{k})
\vec{F}^i \mathbf{\Psi}\mathbf{\Psi}^\dagger \vec{F}^j
\end{align}
\begin{align}
\nonumber
N_{\vec{k}}=&g_0\rho(0)C\mathbf{\Psi}\mathbf{\Psi}^T
+g_s\rho(0)C\mathbf{\Psi}^T\mathbf{\Psi}\\
&+g_d\rho(0)C\bar{h}^{ij}_{2D}(\vec{k})
\vec{F}^i \mathbf{\Psi}\mathbf{\Psi}^T \vec{F}^j
\end{align}
with $C=1/\sqrt{2}$ and
$\bar{h}^{\perp}_{2D}(\vec{k})$ by Eq. \ref{eq:dipolar_effective}.

\begin{acknowledgments}
We acknowledge useful discussions with 
D. Stamper-Kurn, M. Vengalatorre, A. Lamacraft, V. Gritsev, G. Shlyapnikov.
This work was supported by NDSEG and NSF Graduate Research
Fellowship, Harvard-MIT CUA, DARPA, MURI, and the NSF grant DMR-0705472.
\end{acknowledgments}

\bibliographystyle{pnas-bolker}
\bibliography{dipolar}


\end{document}